\documentclass{article}
\usepackage{mathtools}
\usepackage{graphicx}
\usepackage{amsmath}
\usepackage{amssymb}
\usepackage{amsthm}
\usepackage{iftex}
\usepackage{lmodern}
\usepackage{textcomp}
\usepackage{multirow}
\usepackage[symbol]{footmisc}

\theoremstyle{thmstyleone}
\newtheorem{theorem}{Theorem}[section]

\newtheorem{example}{Example}[section]%
\newtheorem{lemma}{Lemma}[section]%

\begin{document}

\begin{center}
	\textbf{\huge Reversible complement cyclic codes over finite chain rings}
\end{center}
\begin{center}
	Monika Dalal$^{1}$, Sucheta Dutt$^{1*}$ and Ranjeet Sehmi$^{1}$\\
	$^{1}$Department of Applied Sciences,Punjab Engineering College\\
	(Deemed to be University), Chandigarh, India, 160012\\
	$^{*}$Corresponding author(s). E-mail(s): $rsehmi@pec.edu.in$ ;\\
	Contributing authors : $monika.phdappsc@pec.edu.in$;\\
	$sucheta@pec.edu.in$; 
\end{center}

\section*{Abstract}\label{sec0}

       Let $k$ be an arbitrary element of a finite commutative chain ring $R$ and $u$ be a unit in $R.$ In this work, we present necessary conditions which are sufficient as well for  a cyclic code to be a $(u,k)$ reversible complement code over $R.$ Using these conditions, all principally generated cyclic codes over the ring $Z_{_{2}}+vZ_{_{2}}+v^{2}Z_{_{2}}, v^{3}=0$  of length $4$ have been checked to find whether they are $(1,1)$ reversible complement or not.\\\\
        
       \textbf{Keywords :} Code, Cyclic, Reversible, Complement \\

\section{Introduction}\label{sec1}

       The class of cyclic codes is a crucial class of linear block codes where cyclic shifts of every codeword belong to the code itself. Reversible and reversible complement cyclic codes are important subclasses of cyclic codes which find applications in DNA data storage, molecular computation, quantum computing, cryptography and communication networks. A vast literature is available on reversible cyclic codes over fields, Galois rings and finite chain rings \cite{1,2,3,4,5}.  Reversible complement cyclic codes over different rings have also been explored by various researchers \cite{6,7,8,9,10,11,12,13}. J. Kaur et al. have defined a generalized notion for complement of an element of a finite commutative ring in \cite{6}. They have used this notion to derive necessary conditions for a cyclic code over a Galois ring to be a  reversible complement, which are sufficient as well. 
       
       In this work, we present necessary conditions which are sufficient as well for a cyclic code to be a $(u,k)$ reversible complement code over $R,$ thereby extending the results of \cite{6}. Using these conditions, all principally generated cyclic codes over the ring $Z_{_{2}}+vZ_{_{2}}+v^{2}Z_{_{2}}, v^{3}=0$ of length $4$ have been checked to find whether they are $(1,1)$ reversible complement or not.

\section{Preliminaries}

      Let $R$ be a finite commutative ring having unity. If all the ideals of $R$ form a chain under inclusion, then it is said to be a finite commutative chain ring. Let $k$ be an arbitrary element in $R$ and $u$ be a unit in $R$ such that $u^{2}=1$ and $uk=k.$  J. Kaur et al. have shown that for every $r$ belonging to $R,$ an element $\overline{r}$ belonging to $R$ can be found such that $r+u\overline{r}=k$ \cite{6}. It is easily seen that $\overline{(\overline{r})}=r.$ This elememt $\overline{r}$ of $R$ is defined as the complement of $r$ with respect to $u$ and $k,$ by J. Kaur et al. Henceforth, we shall denote $\overline{r}$ by $(r)^{c}_{_{(u,k)}}$ to emphasize its dependency on $u$ and $k$ and call it the $(u,k)$ complement of $r.$ % The $(u,k)$ reverse complement of a polynomial $a(z)=a_{_{0}}+a_{_{1}}z+\cdots +a_{_{s}}z^{s}$  of degree $s$ over $R$ is defined by $\big( a(z) \big)^{RC}_{_{(u,k)}} = (a_{_{s}})^{c}_{_{(u,k)}}+(a_{_{s-1}})^{c}_{_{(u,k)}}z+\cdots +(a_{_{0}})^{c}_{_{(u,k)}}z^{s}.$
      A linear code $C$ over $R$ is said to be a cyclic code if for every codeword in $C$, all of its cyclic shifts also belong to $C.$ A cyclic code $C$ having length $n$ over a ring $R$ is said to be a reversible cyclic code if $(a_{_{n-1}},a_{_{n-2}},\cdots,a_{_{0}}) \in C$ for every $(a_{_{0}},a_{_{1}},\cdots,a_{_{n-1}}) \in C$. Let $a(z)$ be equal to $ a_{_{0}}+a_{_{1}}z+\cdots +a_{_{n-1}}z^{n-1}$ be the polynomial representation of the codeword $(a_{_{0}},a_{_{1}}, \cdots, a_{_{n-1}}).$ The reciprocal polynomial of a polynomial $a(z)$ in $C$ is denoted by $a^{*}(z)$ and is equal to  $z^{deg\big(a(z)\big)} a(1/z).$ The $(u,k)$ reverse complement of $a(z)$ is defined by $(a_{_{n-1}})^{c}_{_{(u,k)}}+(a_{_{n-2}})^{c}_{_{(u,k)}}z+\cdots +(a_{_{0}})^{c}_{_{(u,k)}}z^{n-1}$ and is denoted by $\big( a(z) \big)^{RC}_{_{(u,k)}}.$ A cyclic code $C$  is called a $(u,k)$ reversible complement code if $\big(a(z)\big)^{RC}_{_{(u,k)}} \in C$ for every $a(z) \in C.$

\section{Reversible complement cyclic codes}

      In this section, let $R$ denote a finite chain ring and $C$ denote a cyclic code having arbitrary length $n$ over $R.$  We obtain necessary conditions which are sufficient as well for $C$ to be a $(u,k)$ reversible complement code using the reversiblity conditions established by Monika et al. in \cite{5}. We also present some examples of reversible complement cyclic codes over such rings. For the sake of completeness, we recall the structure of $C$ over $R$ \cite{14} and the conditions for reversibility of $C$ given by Monika et al. \cite{5}.
      
      Let $S=\{f_{_{0}}(z),f_{_{1}}(z), \cdots, f_{_{m}}(z)\}$ be a set of minimal degree polynomials of certain subsets of $C$ with leading coefficient of $f_{_{j}}(z)$ equal to $\gamma^{i_{_{j}}}u_{_{j}},$ where $u_{_{j}}$ is some unit in $R,$ $deg\big(f_{_{j}}(z) \big) < deg\big( f_{_{j+1}}(z) \big),$ $i_{_{j}}>i_{_{j+1}}$ and $i_{_{j}}$ is the smallest such power. If $i_{_{0}}=0,$ then $f_{_{0}}(z)$ is monic and we have $m=0.$

      \begin{theorem} (\cite{14})
      	Suppose $C$ is a cyclic code over $R$ and $S=\{f_{_{0}}(z),f_{_{1}}(z), \cdots, f_{_{m}}(z)\},$ where $f_{_{j}}(z)$ are polynomials of $C$ as defined above. Then
      	\begin{itemize}
      	\item[$(i)$] $S$ is a generating set for $C.$
      	\item[$(ii)$]For every $j,$ $0 \leq j \leq m$, $f_{_{j}}(z) = \gamma^{i_{_{j}}} h_{_{j}}(z)$  such that $h_{_{j}}(z)$ is a monic polynomial over the finite chain ring having nilpotency index $\nu - i_{_{j}}$ and maximal ideal $\langle \gamma\rangle$.
      	\end{itemize}
      \end{theorem}

      \begin{theorem} (\cite{5})
      	Let $C$ be a cyclic code over $R$ generated by $S$ as defined earlier. Then $C$ is reversible if and only if
      	\begin{itemize}
      	\item[$(i)$] $f^{*}_{_{0}}(z)=u_{_{0}}f_{_{0}}(z)$ for some unit $u_{_{0}} \in R,$
      	\item[$(ii)$] $f^{*}_{_{r}}(z)-u_{_{r}}f_{_{r}}(z) \in \langle f_{_{s}}(z),f_{_{s-1}}(z), \cdots, f_{_{0}}(z)\rangle$ for some $s<r$ and a unit $u_{_{r}} \in R$, $0<r \leq \nu-1$.
      \end{itemize}
      \end{theorem}

      We require the following lemma $3.1$ to establish the main result of this article.
      \begin{lemma}
      	Let $C$ be a cyclic code having length $n$ over $R$ generated by the set $S.$ Let $a(z)$ be an arbitrary polynomial of degree $s$ in $C$ and $0(z)$ denote the zero polynomial in $C.$ Then, the $(u,k)$ reverse complement of $a(z)$ can be expressed as %in a combination of the reciprocal polynomial of $a(z)$ and the reverse complement of the zero polynomial of $C$ as 
      	$$\big( a(z) \big)^{RC}_{_{(u,k)}} = \big( 0(z) \big)^{RC}_{_{(u,k)}} - u^{-1}z^{n-s-1}a^{*}(z),$$
      	where $a^{*}(z)$ is the reciprocal polynomial of $a(z).$
      \end{lemma}

      \begin{proof}
      	Let $0(z)$ be the zero polynomial of $C$ and $a(z) = a_{_{0}}+a_{_{1}}z+\cdots + a_{_{s}}z^{s},$ for $a_{_{i}} \in R,$ $0 \leq i \leq s < n$ be an arbitrary polynomial of degree $s$ in $C.$ Then $\big( 0(z) \big)^{RC}_{_{(u,k)}} = u^{-1}k(z^{n-1}+z^{n-2}+\cdots +z+1) $ and $\big( a(z) \big)^{RC}_{_{(u,k)}} = (a_{_{0}})^{c}_{_{(u,k)}}z^{n-1} +(a_{_{1}})^{c}_{_{(u,k)}}z^{n-2}+\cdots + (a_{_{s}})^{c}_{_{(u,k)}}z^{n-s-1} +u^{-1}k(z^{n-s-2}+\cdots +z+1).$ Therefore, $\big( 0(z) \big)^{RC}_{_{(u,k)}}-\big( a(z) \big)^{RC}_{_{(u,k)}} = \big(u^{-1}k-(a_{_{0}})^{c}_{_{(u,k)}}\big)z^{n-1}+\big(u^{-1}k-(a_{_{1}})^{c}_{_{(u,k)}}\big)z^{n-2}+\cdots +\big(u^{-1}k-(a_{_{s}})^{c}_{_{(u,k)}}\big)z^{n-s-1} = u^{-1}\big( a_{_{0}}z^{n-1}+a_{_{1}}z^{n-2}+\cdots+a_{_{s}}z^{n-s-1} \big) = u^{-1}z^{n-s-1}a^{*}(z).$ Thus, $\big( a(z) \big)^{RC}_{_{(u,k)}} = \big( 0(z) \big)^{RC}_{_{(u,k)}} - u^{-1}z^{n-s-1}a^{*}(z).$
      \end{proof}

      \begin{theorem}
      	Consider a cyclic code $C$ having length $n$ over $R$, generated by $S$ as defined earlier. Then, $C$ is a $(u,k)$ reversible complement code over $R$ if and only if 
      	\begin{itemize}
     		\item[$(i)$] $\big( 0(z) \big)^{RC}_{_{(u,k)}} \in C,$
     		\item[$(ii)$]$ f^{*}_{_{0}}(z)=u_{_{0}}f_{_{0}}(z)$ for some unit $u_{_{0}} \in R,$
     		\item[$(iii)$] $f^{*}_{_{r}}(z)-u_{_{r}}f_{_{r}}(z) \in \langle f_{_{s}}(z),f_{_{s-1}}(z), \cdots, f_{_{0}}(z)\rangle$ for some $s<r$ and a unit $u_{_{r}} \in R$, $0<r \leq \nu-1$.
      	\end{itemize}      
      \end{theorem}

      \begin{proof}
      	Let $C$ be a cyclic code having length $n$ over $R$ such that $C$ is a $(u,k)$ reversible complement code. Then $\big( a(z) \big)^{RC}_{_{(u,k)}} \in C$ for every $a(z)$ belonging to $C.$ In particular, $\big( 0(z) \big)^{RC}_{_{(u,k)}} \in C.$ This proves $(i).$ Using Lemma 3.1, we get that $ u^{-1}z^{n-s-1}a^{*}(z) = \big( 0(z) \big)^{RC}_{_{(u,k)}}-\big( a(z) \big)^{RC}_{_{(u,k)}} \in C.$ Since $C$ is cyclic, it follows that $a^{*}(z) \in C$ for every $a(z) \in C.$ Thus, $C$ is reversible. Conditions $(ii)$ and $(iii)$ now follow from Theorem 3.2. 
      	
      	Conversely, let $C$ be a cyclic code having length $n$ over $R$ such that $(i),$ $(ii)$ and $(iii)$ hold. Conditions $(ii)$ and $(iii)$ imply that $C$  is reversible  by Theorem 3.2 and therefore, $a^{*}(z)$ belongs to $C$ for every $a(z)$ belonging to $C.$ This together with  condition $(i)$ and Lemma 3.1 implies that $\big( a(z) \big)^{RC}_{_{(u,k)}} \in C$ for every $a(z)$ belonging to $C.$ Thus, $C$ is a $(u,k)$ reversible complement code over $R.$
      \end{proof}

       Next, we provide some examples to support this result.

       \begin{example}
       	Let $C$ be a cyclic code with length 4 over the ring $R=Z_{_{2}}+vZ_{_{2}}+v^{2}Z_{_{2}}, v^{3}=0$ of $characteristic$ 2. The following table classifies all principally generated cyclic codes of length 4 over $R$ into $(1,1)$ reversible complement cyclic codes and not $(1,1)$ reversible complement cyclic codes. We shall denote $z+1$ by $h$ in this table.
       \end{example}
       	
       	\begin{center}

       	\begin{tabular}{|c|c|c|}
       		\hline 
       		S. No. & Cyclic code C  & $(1,1)$ Reversible complement  \\ 
       		\hline 
       		1 & $\langle 0 \rangle$ & No \\ 
       		\hline 
       		2 & $\langle v^{2} \rangle$ & No \\ 
       		\hline 
       		3 & $\langle v^{2}h \rangle$ & No \\ 
       		\hline 
       		4 & $\langle v^{2}h^{2} \rangle$ & No \\ 
       		\hline 
       		5 & $\langle v^{2}h^{3} \rangle$ & No \\ 
       		\hline 
       		6 & $\langle v \rangle$ & No \\ 
       		\hline 
       		7 & $\langle vh \rangle$ & No \\ 
       		\hline 
       		8 & $\langle vh^{2} \rangle$ & No \\ 
       		\hline 
       		9 & $\langle vh^{3} \rangle$ & No \\ 
       		\hline 
       		10 &$ \langle 1 \rangle$ & Yes \\ 
       		\hline 
       		11 & $\langle h \rangle$ & Yes \\ 
       		\hline 
       		12 & $\langle h^{2} \rangle$ & Yes  \\ 
       		\hline 
       		13 & $\langle h^{3} \rangle$ & Yes \\ 
       		\hline 
       		14 & $\langle vh+v^{2} \rangle$ & No \\ 
       		\hline 
       		15 & $\langle vh^{2}+v^{2} \rangle$ & No \\ 
       		\hline 
       		16 & $\langle vh^{2}+v^{2}h \rangle$ & No \\ 
       		\hline 
       		17 & $\langle vh^{2}+v^{2}\big( h+1 \big) \rangle$ & No \\ 
       		\hline 
       		18 & $\langle vh^{3}+v^{2} \rangle$ & No \\ 
       		\hline 
       		19 & $\langle vh^{3}+v^{2}h(h+1) \rangle$ & No \\ 
       		\hline 
       		20 & $\langle vh^{3}+v^{2}h \rangle$ & No \\ 
       		\hline 
       		21 & $\langle h+v^{2} \rangle$ & Yes \\ 
       		\hline 
       		22 & $\langle h^{2}+v^{2} \rangle$ & No \\ 
       		\hline 
       		23 & $\langle h^{3}+v^{2} \rangle$ & No \\ 
       		\hline
       		24 & $\langle h^{3}+v^{2}h \rangle$ & No \\ 
       		\hline 
       		25 & $\langle h^{3}+v^{2}(h+1) \rangle$ & No \\ 
       		\hline   
       	\end{tabular} 
        \end{center}
    
        \begin{center}
       	\begin{tabular}{|c|c|c|}
       	\hline 
       	S. No. & Cyclic code C  & $(1,1)$ Reversible \\
       	 & & complement  \\ 
       	\hline
       	26 & $\langle h+v \rangle$ & Yes \\ 
       	\hline
       	27 & $\langle h^{2}+v(h+1) \rangle$ & No \\ 
       	\hline 
       	28 & $\langle h^{2}+vh \rangle$ & No \\ 
       	\hline 
       	29 & $\langle h^{2}+v \rangle$ & No \\ 
       	\hline  
       	30 & $\langle h^{3}+v \rangle$ & No \\ 
       	\hline 
       	31 & $\langle h^{3}+vh(h+1) \rangle$ & No \\ 
       	\hline 
       	32 & $\langle h^{3}+vh \rangle$ & No \\ 
       	\hline 
       	33 & $\langle h+v+v^{2} \rangle$ & Yes \\ 
       	\hline 
       	34 &$ \langle h^{2}+v(h+1)+v^{2}(h+1) \rangle$ & No \\ 
       	\hline 
       	35 & $\langle h^{2}+v(h+1)+v^{2} \rangle$ & No \\ 
       	\hline 
       	36 & $\langle h^{2}+v(h+1)+v^{2}h \rangle$ & No  \\ 
       	\hline 
       	37 & $\langle h^{2}+v+v^{2} \rangle$ & No \\ 
       	\hline 
       	38 & $\langle h^{2}+vh+v^{2}(h+1) \rangle$ & Yes \\ 
       	\hline 
       	39 & $\langle h^{2}+vh+v^{2} \rangle$ & Yes \\ 
       	\hline 
       	40 & $\langle h^{2}+vh+v^{2}h \rangle$ & No \\ 
       	\hline 
       	41 & $\langle h^{3}+vh(h+1)+v^{2} \rangle$ & No \\ 
       	\hline 
       	42 & $\langle h^{3}+vh(h+1)+v^{2}h \rangle$ & No \\ 
       	\hline 
       	43 & $\langle h^{3}+vh(h+1)+v^{2}h^{2} \rangle$ & No \\ 
       	\hline 
       	44 & $\langle h^{3}+vh+v^{2} \rangle$ & No \\ 
       	\hline 
       	45 & $\langle h^{3}+vh+v^{2}\big( h^{2}+h+1 \big) \rangle$ & No \\ 
       	\hline 
       	46 & $\langle h^{3}+vh+v^{2}(h+1) \rangle$ & No \\ 
       	\hline  
       \end{tabular}
       \end{center}

\section{Conclusion}
        
       In this article, we have presented necessary conditions which are sufficient as well for a cyclic code to be a $(u,k)$ reversible complement code over a finite chain ring $R$. Using these conditions, we have checked all principally generated cyclic codes with length $4$ over the ring $Z_{_{2}}+vZ_{_{2}}+v^{2}Z_{_{2}}, v^{3}=0$ to find whether they are  $(1,1)$ reversible complement codes over this ring or not.

\section*{Acknowledgements}

       The first author gratefully acknowledges the support provided by the Council of Scientific and Industrial Research (CSIR), India in the form of a research fellowship.

%\section*{Declarations}

\end{document}